\documentclass[twocolumn, pra, aps]{revtex4}

\usepackage{amsmath}
\usepackage{amsfonts}
\usepackage{bm}
\usepackage{graphicx}

\begin{document}

\preprint{APS/123-QED}

\title{$n$-body Correlation of Tonks-Girardeau Gas}
\author{Yajiang Hao}
\email{haoyj@ustb.edu.cn}
\affiliation{Institute of Theoretical Physics and Department of Physics, University of Science and Technology Beijing, Beijing 100083, China}
\author{Yaling Zhang}
\affiliation{Institute of Theoretical Physics and Department of Physics, University of Science and Technology Beijing, Beijing 100083, China}
\author{Yiwang Liu}
\affiliation{Institute of Theoretical Physics and Department of Physics, University of Science and Technology Beijing, Beijing 100083, China}

\author{Li Wang}
\affiliation{Institute of Theoretical Physics, State Key Laboratory of Quantum Optics and Quantum Optics Devices, Collaborative Innovation Center of Extreme Optics, Shanxi University, Taiyuan 030006, P. R. China}
\begin{abstract}
For the well-known exponential complexity it is a giant challenge to calculate the correlation function for general many-body wave function. We investigate the ground state $n$th-order correlation functions of the Tonks-Girardeau (TG) gases. Basing on the wavefunction of free fermions and Bose-Fermi mapping method we obtain the exact ground state wavefunction of TG gases. Utilizing the properties of Vandermonde determinant and Toeplitz matrix, the $n$th-order correlation function is formulated as $(N-n)$-order Toeplitz determinant, whose element is the integral dependent on 2$(N-n)$ sign functions and can be computed analytically. By reducing the integral on domain $[0,2\pi]$ into the summation of the integral on several independent domains, we obtain the explicit form of the Toeplitz matrix element ultimately. As the applications we deduce the concise formula of the reduced two-body density matrix and discuss its properties. The corresponding natural orbitals and their occupation distribution are plotted. Furthermore, we give a concise formula of the reduced three-body density matrix and discuss its properties. It is shown that in the successive second measurements, atoms appear in the regions where atoms populate with the maximum probability in the first measurement.
\end{abstract}

\date{\today}

\maketitle

\section{Introduction}
Since Glauber generalized the first-order correlation to higher-order correlation in the optical research \cite{Glauber}, $n$-body correlation has gradually become one of the foundational properties of many-body quantum systems, which is extremely important not only to the definition of coherence but also to the characterization of the properties of quantum matter including the quantum phases and topological states. Quantum correlation is also the crucial resource for quantum information and computation \cite{ABergschneider2019}. The study of correlation has become the driving force for the development of many research fields. For example, the famous Hanbury Brown and Twiss experiment \cite{HBT,MSchellekens,TJeltes} has ever kept pushing the development of quantum optics. The calculation of correlation functions played a pivotal role in the theoretical study of many-body quantum system. It is helpful to reveal and facilitate the understanding of those exotic quantum effects. For most systems the analytical calculation of correlation functions remains intractable, particularly the higher-order correlation functions, although the high-order correlation is required by rigorous description of the coherence. Experimentally, with the development of cold atom technique including the single-atom-sensitive detection techniques, the third-order \cite{SSHodgman2011,PMPreiss}, the fourth order \cite{VGuarrera} , and even sixth-order \cite{RGDall,GHerce} correlation of ultracold Bose atoms is measurable. With the development of the quantum gas microscopes \cite{PMPreiss,ABergschneider2018,MHolten}, the momentum microscope \cite{SSHodgman2017} and quantum ghost imaging technique \cite{SSHodgman2019} the measurement of the high-order correlation of quantum many-body system has become feasible.

To develop the techniques that measure and control the quantum phase as well as coherence is one of the most important goals of cold atom research, both of which are closely related with the quantum correlation. The application of optical lattice and Feshbach resonance technique extremely improves the controllability of the dimension \cite{NJVDruten,BParedes,TKinoshita} and interacting regime \cite{BParedes,TKinoshita,TJacqmin} of cold atom system. The great progress in experiment has made it a popular platform to investigate the basic problems of quantum many-body system. One of the remarkable achievements of the above techniques is the experimental realization of the strong correlated Tonks-Girardeau (TG) gas \cite{LTonks,MDGirardeau1960},  a one-dimensional neutral Bose atom gas with infinitely strong repulsive interaction. TG gas has now constituted one important portion of the low dimensional quantum gas research \cite{RMP2011,RMP2012,RMP2013}.

Theoretically the TG gas was first studied as a \emph{toy model}  \cite{LTonks,MDGirardeau1960}, its eigen wavefunction can be exactly obtained based on the many-particle wavefunction of polarized fermions utilizing the Bose-Fermi mapping method. Since its experimental realization the static and dynamical structure factor\cite{EJKPN,NFabbri}, universal contact \cite{WXu}, noise correlation \cite{LMathey,KHe,VGritsev}, full counting statistics \cite{PDevillard2020} have been studied. Although the one-body, two-body and local three-body correlation functions \cite{PDevillard2021,ATenart,Olshanii2017}, momentum distribution \cite{ALenard,PJForrester, ILovas2017A,ILovas2017B} and its dynamics \cite{VIYukalov,RPezer,JDobrzyniecki} have ever been studied, so far the explicit formula for the $n$th-order correlation function of TG gases is still lacking.

The motivation of the present paper is to analytically derive a concise formula of the $n$th-order correlation function for TG gas. It is important for the accurate understanding of the many-body quantum correlation in the future reachable experiments. The paper is organized as follows. In Sec. II, we give a brief review of Bose-Fermi mapping method and introduce the ground state wavefunction for TG gases. In Sec. III, we present the general explicit expression of the $n$-body correlation function. In Sec. IV, the reduced two-body density matrix and the reduced three-body density matrix are investigated as examples. A brief summary is given in Sec. V.

\section{Ground state wavefunction}
As the temperature is low enough and the transversal confinement is strong enough to froze the motion of atoms in transversal direction, the quantum gas can only distribute in the longitudinal directions. The system composed of $N$ cold atoms with mass $m$ can be described by the Hamiltonian
\begin{equation}
H=\sum_{j=1}^N-\frac{\hbar ^2}{2m}\frac{\partial ^2}{\partial x_i^2}+g_{1D}\sum_{j<l}\delta(x_j-x_l),
\end{equation}
where atoms interact by $\delta$-potential and the effective interaction strength $g_{1D}$ can be tuned to infinite strong repulsive interaction by Feshbach resonance and confinment induced resonance such that the cold atoms are in the TG regime. In this situation, the many-body wavefunction $\Psi(x_{1},\cdots ,x_{N})$ satisfies not only the Schr\"{o}dinger equation $H\Psi(x_{1},\cdots ,x_{N})=E\Psi(x_{1},\cdots ,x_{N})$ but also the boundary condition $\Psi(x_{1},\cdots ,x_{N})=0$ for $x_j=x_l$. This condition is equivalent to the constraint on the wavefunction of identical fermions for the Pauli exclusive principle. Therefore we can construct the exact wavefunction of TG gas basing on the wavefunction of noninteracting polarized fermions. In the present paper we will assume atoms distribute in a circular ring with length $L$. With the periodical boundary condition the single particle eigenfunction is formulated as $\phi_j(x)=\frac{1}{\sqrt{L}}e^{ik_jx}$
with $k_{j}=2j\pi /L$ ($j=0,\pm 1,\pm 2, \cdots$). The wavefunction of $N$ fermions is the Slater determinant of $\phi_j(x_n)$
\begin{eqnarray}
\Psi_F (x_{1},\cdots ,x_{N})& =(L^{N}N!)^{-1/2}\det
[e^{ik_{l}x_{j}}]_{l=1,...,N}^{j=1,...,N}
\end{eqnarray}
with $k_{l} =2\pi /L(l-(N+1)/2)$. Here $N$ is assumed to be odd for simplicity. Using the properties of Vandermonde determinant formula
\begin{equation*}
\det [(x_{k})^{j-1}]_{j,k=1,\cdots ,N}=\prod_{1\leq j<k\leq N}(x_{j}-x_{k}),
\end{equation*}
we reformulate the determinant form of Fermi wavefunction as the simplified product form
\begin{eqnarray*}
\Psi_F (x_{1},\cdots ,x_{N})&=(L^{N}N!)^{-1/2}\prod_{j=1,...,N}e^{-i(N-1)\pi x_{j}/L}    \\
&\times \prod_{1\leq j<l\leq N}(e^{i2\pi x_{j}/L}-e^{i2\pi x_{l}/L}).
\end{eqnarray*}
This is important to reduce the computation complexity in the later calculation of correlation functions. With the reduction the computation efficiency reduces to $N$-scaling from the $N^2$-scaling. In order to obtain the exchange symmetrical wavefunction satisfied by the identical bosons the Bose-Fermi mapping method can be utilized with the sign function $\epsilon (x)$, which is 0, +1, or -1 for $x$=0, $>0$ or $<0$, respectively. The ground state wavefunction of TG gas is expressed as \cite{LTonks,MDGirardeau1960}
\begin{eqnarray}
&\Psi\left( x_{1},x_{2},\cdots ,x_{N}\right)
=(L^{N}N!)^{-1/2}\prod_{j=1,...,N}e^{-i(N-1)\pi x_{j}/L}   \nonumber \\
&\times \prod_{1\leq j<l\leq N}\epsilon (x_j-x_l)\left( e^{i2\pi x_{j}/L}-e^{i2\pi x_{l}/L}\right).
\end{eqnarray}
For simplicity we take the length unit $2\pi/L$ in the latter part so the length of circular ring is $2\pi$. The original notation will be preserved. Therefore the wavefunction of ground state for TG gases is
\begin{eqnarray}
&&\Psi\left( x_{1},x_{2},\cdots ,x_{N}\right) \\
&=&(\sqrt{2\pi})^{-N}(N!)^{-1/2}\prod_{j=1,...,N}e^{-i(N-1) x_{j}/2}   \nonumber \\
&&\times \prod_{1\leq j<l\leq N}\epsilon (x_j-x_l)\left( e^{i x_{j}}-e^{i x_{l}}\right).  \notag
\end{eqnarray}

\section{$n$-body correlation function}

The $n$-body correlation function is defined as (in unit of $(\frac{2\pi}{L})^n$)
\begin{eqnarray}
&\rho _{n}(x_{N-n+1},\cdots ,x_{N};x_{N-n+1}^{\prime },\cdots ,x_{N}^{\prime }) \notag    \\
=&\frac{N!}{(N-n)!}\int_{0}^{2\pi }dx_{1}\cdots \int_{0 }^{2\pi }dx_{N-n}\Psi ^{\ast }(x_{1},\cdots,x_{N})
\notag \\
& \times \Psi(x_{1},\cdots ,x_{N-n},x_{N-n+1}^{\prime },\cdots ,x_{N}^{\prime }),
\end{eqnarray}
which is the integral over $N-n$ variables ($x_1,\cdots,x_{N-n}$). The multiple integral is usually difficult to analytically calculate and have to resort to numerical method such as Monte Carlo method. In the case of TG gases, the integral can be calculated analytically and we will obtain its explicit formula in this section. Substituting the wavefunction Eq. (4) into Eq. (5) and using the following result for Toeplitz matrix  \cite{ALenard,PJForrester,TPapenbrock,HGVaidyaJMP,DMGangardt}
\begin{eqnarray}
& \frac{1}{N!}\prod_{l=1}^{N}\int_{0}^{2\pi }dx_{l}g(x_{l})\prod_{1\leq
j<k\leq N}|e^{ix_{j}}-e^{ix_{k}}|^{2} \\
& =\det [\int_{0}^{2\pi }dtg(t)\exp (it(j-k))]_{j,k=1,...,N}, \notag
\end{eqnarray}
the above ($N-n$)-fold integral transforms into a ($N-n$)-order Toeplitz determinant
\begin{eqnarray*}
&& \rho _{n}(x_{N-n+1},\cdots ,x_{N};x_{N-n+1}^{\prime },\cdots ,x_{N}^{\prime }) \\
& =&\frac{1}{(2\pi )^{N}}\prod_{j=N-n+1,...,N}e^{i(N-1)\left(
x_{j}-x_{j}^{\prime}\right) /2} \\
&& \prod_{N-n+1\leq j<l\leq N}\epsilon
(x_{j}-x_{l})(e^{-ix_{j}}-e^{-ix_{l}})  \\
&&\times \epsilon(x_{j}^{\prime}-x_{l}^{\prime})(e^{ix_{j}^{\prime}}-e^{ix_{l}^{\prime}}) \\
&& \times \det [b_{j-k}(x_{N-n+1},\cdots ,x_{N};x_{N-n+1}^{\prime },\cdots ,x_{N}^{\prime })]_{j,k=1,...,N-n}
\end{eqnarray*}
with the Toeplitz matrix elements being single-variable integral as follows
\begin{eqnarray*}
&& b_{j-k}(x_{N-n+1},\cdots ,x_{N};x_{N-n+1}^{\prime },\cdots ,x_{N}^{\prime }) \\
&=& \int_{0}^{2\pi }dt\prod_{N-n+1\leq l\leq N}\epsilon (t-x_{l})\epsilon
(t-x_{l}^{\prime})(e^{-it}-e^{-ix_{l}})      \\
&&\times (e^{it}-e^{ix_{l}^{\prime}})\exp (it(j-k)).
\end{eqnarray*}
To simplify the integral we reformulate the above formula as
\begin{eqnarray*}
&& b_{j-k}(x_{N-n+1},\cdots ,x_{N};x_{N-n+1}^{\prime },\cdots ,x_{N}^{\prime }) \\
&=& \int_{0}^{2\pi }dt\prod_{N-n+1\leq l\leq N}\epsilon (t-x_{l})\epsilon
(t-x_{l}^{\prime }) \\
&& \times f(x_{N-n+1},\cdots ,x_{N};x_{N-n+1}^{\prime },\cdots ,x_{N}^{\prime },t),
\end{eqnarray*}
in which
\begin{eqnarray*}
&& f(x_{N-n+1},\cdots ,x_{N};x_{N-n+1}^{\prime },\cdots ,x_{N}^{\prime },t) \\
& =&e^{-i\sum_{l=N-n+1}^{N}x_{l}}\sum_{l=0}^{2n}(-1)^{l+n}c_{l}e^{i(2n-l)t}\exp (it(j-k-n)).
\end{eqnarray*}%
Here $c_{0}=1$ and $c_{l}=\sum_{S_{l}}\exp \left[ i\sum_{j}s_{j}\right] $ ($l=1,\cdots,2n$)
with $S_{l}=\left\{ s_{1},\cdots ,s_{l}\right\} $ enumerating the subsets comprising $l$ elements of $\left\{ x_{N-n+1},\cdots ,x_{N};x_{N-n+1}^{\prime },\cdots ,x_{N}^{\prime
}\right\} $.
In the above integral the $2n$ sign functions make the integral become complicated. The issue can be treated by observing that in the integrand function $x_l$ and $x_l^{\prime}$ ($N-n+1 \leq l\leq N$) has no specific order so we can order them to split the whole integral interval $[0,2\pi]$ into $n$ independent integral interval in which the sign functions take the definite value of 0, 1, or -1. By sorting $x_{N-n+1},\cdots
,x_{N};x_{N-n+1}^{\prime },\cdots ,x_{N}^{\prime }$ in the ascending order
and denoting them as $r_{j}$ ($j=1,2,\cdots ,2n$), we have%
\begin{eqnarray*}
&&b_{j-k}(x_{N-n+1},\cdots ,x_{N};x_{N-n+1}^{\prime },\cdots ,x_{N}^{\prime
}) \\
&=&\left[ \int_{0}^{2\pi }dt-2\left(\int_{r_{1}}^{r_{2}}dt+\int_{r_{3}}^{r_{4}}dt+\cdots
+\int_{r_{2n-1}}^{r_{2n}}dt\right) \right]  \\
&&\times f(x_{N-n+1},\cdots,x_{N};x_{N-n+1}^{\prime },\cdots ,x_{N}^{\prime },t).
\end{eqnarray*}
Finally the Toeplitz matrix elements $b_{j-k}(x_{N-n+1},\cdots ,x_{N};x_{N-n+1}^{\prime },\cdots ,x_{N}^{\prime })$ take the explicit formula
\begin{eqnarray*}
&&b_{j-k}(x_{N-n+1},\cdots ,x_{N};x_{N-n+1}^{\prime },\cdots ,x_{N}^{\prime
}) \\
&=&e^{-i\sum_{l=N-n+1}^{N}x_{l}}\sum_{l=0}^{2n}(-1)^{l+n}c_{l} \\
&& \times\left[ \nu_{j-k-l+n} -2\sum\nolimits_{s=0}^{n-1}\mu_{j-k-l+n}(r_{2s+1},r_{2s+2}) \right]
\end{eqnarray*}
with $\nu _{m}=2\pi \delta _{m,0}$ and $\mu _{m}(p,q)=q-p$ (for $m=0$), or $\frac{1}{im}(e^{imq}-e^{imp})$ (for $m\neq 0$).

For $x_j=x_j^{\prime}$ ($j=N-n+1,\cdots,x_N$), the $n$-body correlation mean the joint probability observing $n$ particles at $x_{N-n+1}$, $x_{N-n+2}$, $\cdots $, and $x_{N}$, respectively. The above formula can be simplified as
\begin{eqnarray*}
&& \rho _{n}(x_{N-n+1},\cdots ,x_{N};x_{N-n+1},\cdots ,x_{N}) \\
& =&\frac{1}{(2\pi )^{N}}\prod_{N-n+1\leq j<l\leq N}\left\vert
e^{-ix_{j}}-e^{-ix_{l}}\right\vert ^{2} \\ &&\times \det [b_{j-k}(x_{N-n+1},\cdots ,x_{N};x_{N-n+1},\cdots
,x_{N})]_{j,k=1,...,N-n}
\end{eqnarray*}
with
\begin{eqnarray*}
&&b_{j-k}(x_{N-n+1},\cdots ,x_{N};x_{N-n+1},\cdots ,x_{N}) \\
&=&2\pi e^{-i\sum_{l=N-n+1}^{N}x_{l}}(-1)^{j-k+2n}c_{j-k+n}.
\end{eqnarray*}

\begin{figure}[htbp]
    \includegraphics[width=3.0in]{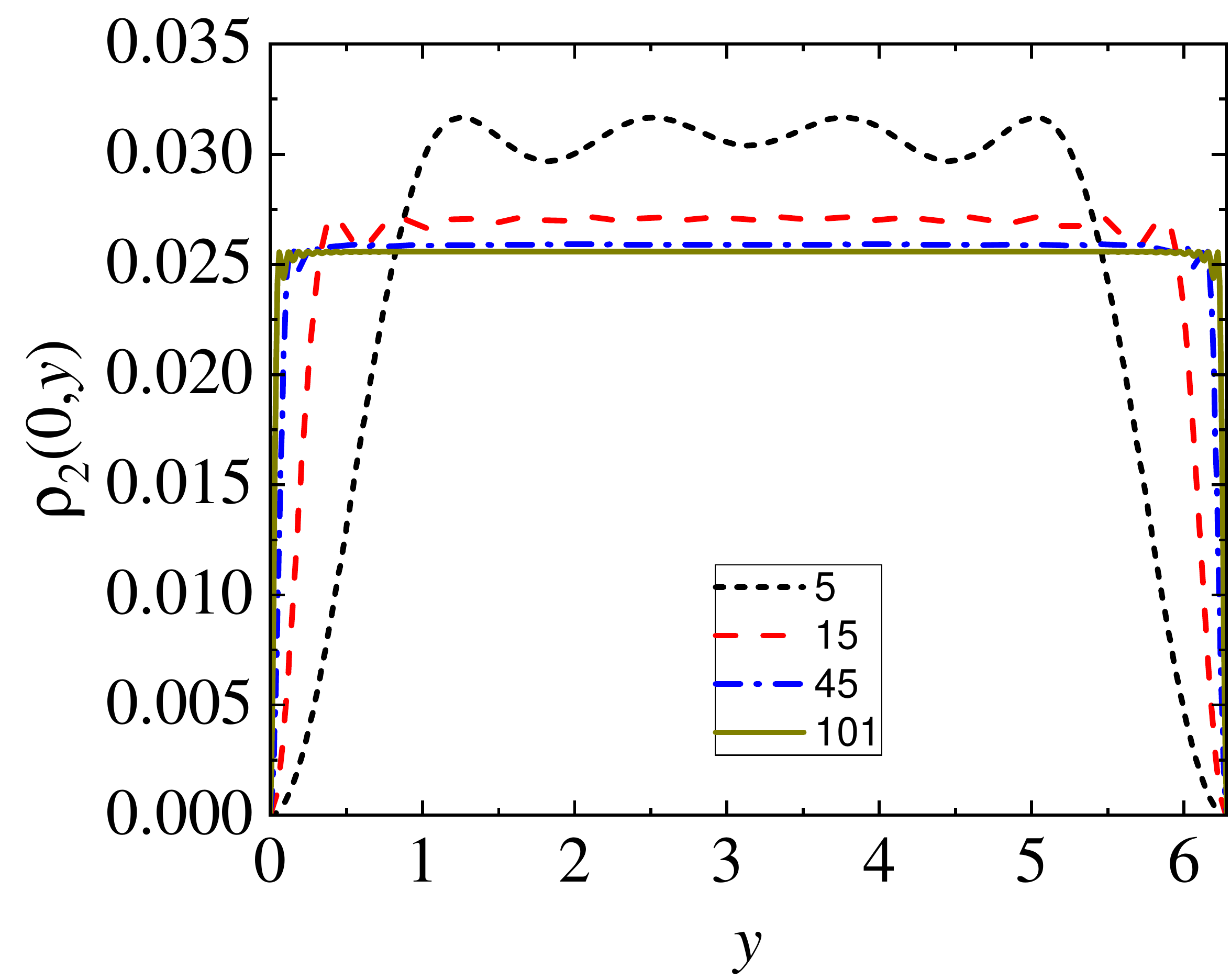}
   \caption{The pair correlation of TG gases. In the figure $\rho_2(0,y)$ has been divided by $N(N-1)$. The atom number is $N=5,15,45,101$ from the above to the below.}
    \label{fig:PC-CS}
\end{figure}
For example, the pair correlation has the following expression
\begin{equation*}
\rho_2\left( x,y\right) =\frac{1}{\pi ^2}\sin^2\frac{y-x}{2}
\det [b_{j-k}\left( x,y\right) ]_{j,k=1,\cdots ,N-2}
\end{equation*}
with
\begin{eqnarray*}
b_{m}\left( x,y\right) &=&2\left( 2+\cos (x-y)\right) \delta _{m,0}-2\left(
e^{ix}+e^{iy}\right) \delta _{m,1} \\
&&-2\left( e^{-ix}+e^{-iy}\right) \delta _{m,-1}+e^{i(x+y)}\delta
_{m,2}  \\
&&+e^{-i(x+y)}\delta _{m,-2}.
\end{eqnarray*}
In the present system satisfying periodical boundary condition the pair correlation is translation invariant, i.e., $\rho_2(x,y)=\rho_2(|x-y|)$. We plot $\rho_2(0,y)$ in Fig. 1 for different atom number. It is shown that for the few atom system the obvious oscillation is displayed while as the atom number is large enough the pair correlation is independent on the atom distance $y$ except the zero distance regime where the pair correlation is zero for the constraint of the infinite repulsion contact boundary condition.

\begin{figure*}[htbp]
    \centering
    \includegraphics[width=7in]{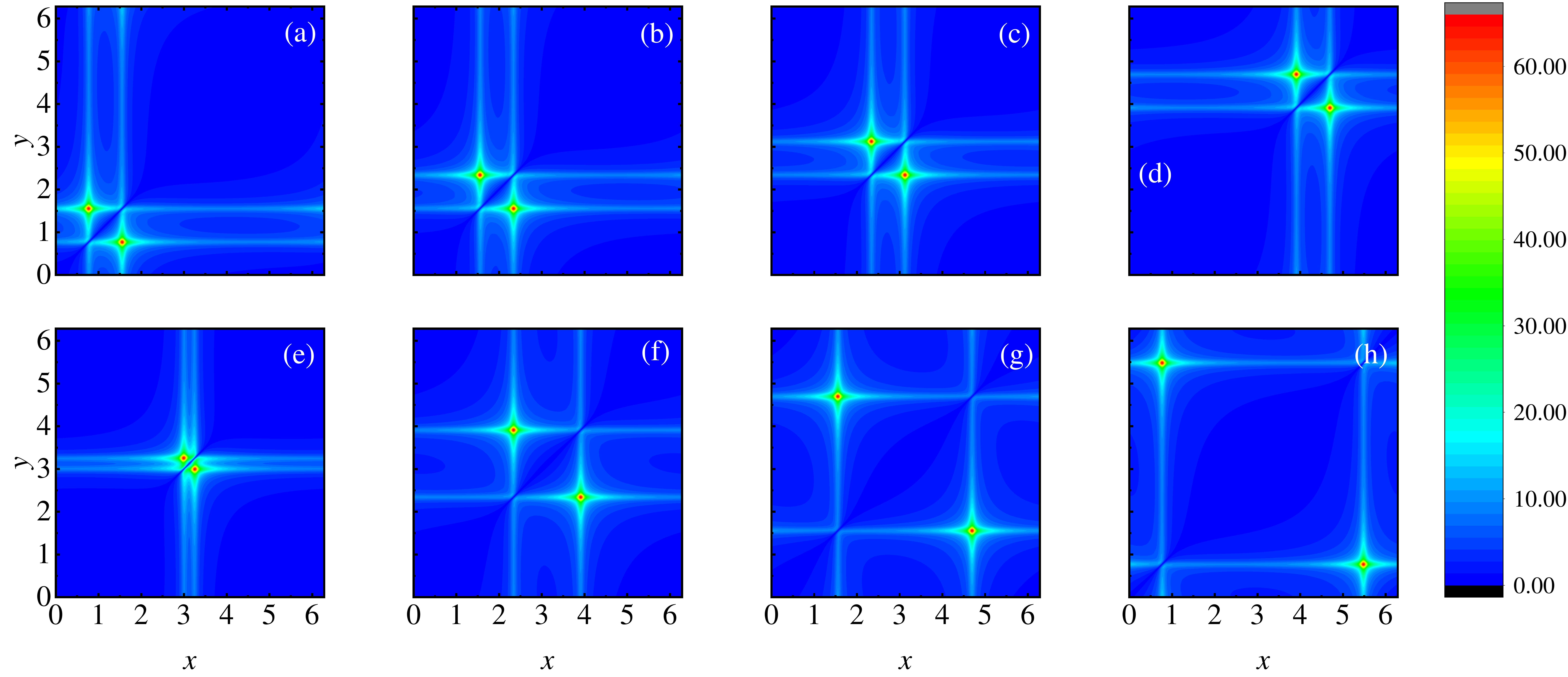}
   \caption{The reduced 2-body density matrix $\rho_2(x,y,x_0^{\prime},y_0^{\prime})$ for TG gas with $N=51$. (a) $x_0^{\prime}=\pi/4,y_0^{\prime}=\pi/2$; (b) $x_0^{\prime}=\pi/2,y_0^{\prime}=3\pi/4$; (c) $x_0^{\prime}=3\pi/4,y_0^{\prime}=\pi$; (d) $x_0^{\prime}=5\pi/4,y_0^{\prime}=3\pi/2$; (e) $x_0^{\prime}=23\pi/24,y_0^{\prime}=25\pi/24$; (f) $x_0^{\prime}=3\pi/4,y_0^{\prime}=5\pi/4$; (g) $x_0^{\prime}=\pi/2,y_0^{\prime}=3\pi/2$; (h) $x_0^{\prime}=\pi/4,y_0^{\prime}=7\pi/4$.}
    \label{fig:R2BDM-CS}
\end{figure*}

\section{The Reduced Two-body and Three-body Density Matrix}

In this section, we show the reduced two-body density matrix (R2BDM) and the reduced three-body density matrix (R3BDM) as examples.

The R2BDM is the probability to find two particles at the positions $x^{\prime}$ and $y^{\prime}$ in the first measurement and  at the positions $x$ and $y$ in the successive second measurements, which has the following concise form
\begin{eqnarray*}
\rho _{2}(x,y;x^{\prime },y^{\prime })& =\frac{1}{(2\pi )^{N}}%
e^{i(N-1)(x+y-x^{\prime }-y^{\prime })/2}\epsilon (x-y) \\
& \times \epsilon (x^{\prime}-y^{\prime })(e^{-ix}-e^{-iy})(e^{ix^{\prime }}-e^{iy^{\prime }}) \\
& \times \det [b_{j-k}(x,y;x^{\prime },y^{\prime })]_{j,k=1,...,N-2}.
\end{eqnarray*}

The matrix element $b_{j-k}(x,y;x^{\prime },y^{\prime })$ can be calculated by (here we sort $x,y,x^{\prime },y^{\prime }$ in the ascending order and denote them as $r_i$ ($i=1,2,3,4$) with $r_1 \leq r_2 \leq r_3 \leq r_4$)
\begin{eqnarray*}
b_{j-k}(x,y;x^{\prime },y^{\prime }) &=&\sum_{l=-2}^{2}(-1)^{l}m_{l}\left[ \nu _{j-k+l} \right.   \\
&& \left.-2\left( \mu_{j-k+l}(r_{1},r_{2}) +\mu _{j-k+l}(r_{3},r_{4})\right) \right]
\end{eqnarray*}
with $m_{2}=e^{-i(x+y)}$, $m_{-2}=e^{i(x^{\prime }+y^{\prime })}$, $m_{1}=e^{-iy}+e^{-ix}+e^{i(y^{\prime}-x-y)}+e^{i(x^{\prime }-x-y)}$, $m_{-1} =e^{iy^{\prime }}+e^{ix^{\prime }}+e^{i(x^{\prime }+y^{\prime}-y)}+e^{i(x^{\prime }+y^{\prime }-x)}$, and $m_{0} =1+e^{i(y^{\prime }-y)}+e^{i(x^{\prime }-y)}+e^{i(y^{\prime}-x)}
+e^{i(x^{\prime }-x)}+e^{i(x^{\prime }+y^{\prime }-x-y)}$.

In Fig. 2 we show the occupation probability in different regimes in the successive measurement for the given atom positions $x_0^{\prime}$ and $y_0^{\prime}$ in the first measurement. The R2BDM $\rho_2(x,y;x_0^{\prime},y_0^{\prime})$ of TG gases for $N=51$ are displayed for different $x_0^{\prime}$ and $y_0^{\prime}$. We take $y_0^{\prime}-x_0^{\prime}=\pi/4$ in Fig. 2a-Fig. 2d, i.e., in the first measurement two atoms locate at different positions with the atom distance being $\pi/4$. It is shown that $\rho_2(x_0^{\prime},y_0^{\prime};x_0^{\prime},y_0^{\prime})$ and $\rho_2(y_0^{\prime},x_0^{\prime};x_0^{\prime},y_0^{\prime})$ have the maximum value, which means that in the second measurement one atom appears at $x_0^{\prime}$ and the other atom appears at $y_0^{\prime}$ with the greatest probability. Besides the regions close to ($x_0^{\prime}$, $y_0^{\prime}$) and ($y_0^{\prime}$, $x_0^{\prime}$), two atoms also populate in the following regions with a certain probability: $(x_0^{\prime},y)$ ($0\leq y\leq 2\pi$ except $y=x_0^{\prime}$), $(y_0^{\prime},y)$ ($0\leq y\leq 2\pi$ except $y=y_0^{\prime}$), $(x,y_0^{\prime})$ ($0\leq x\leq 2\pi$ except $x=y_0^{\prime}$), and $(x,x_0^{\prime})$ ($0\leq x\leq 2\pi$ except $x=x_0^{\prime}$). The cases for increased atom distance $y_0^{\prime}-x_0^{\prime}=\pi/12$, $ \pi/2$, $\pi$, and $3\pi/2$ are plotted in Fig. 2e-Fig. 2h. The same properties as the previous case are displayed, so the correlation properties of TG gases are not related with the atom distance in the first measurement.

\begin{figure}[htbp]
    \centering
    \includegraphics[width=3.5in]{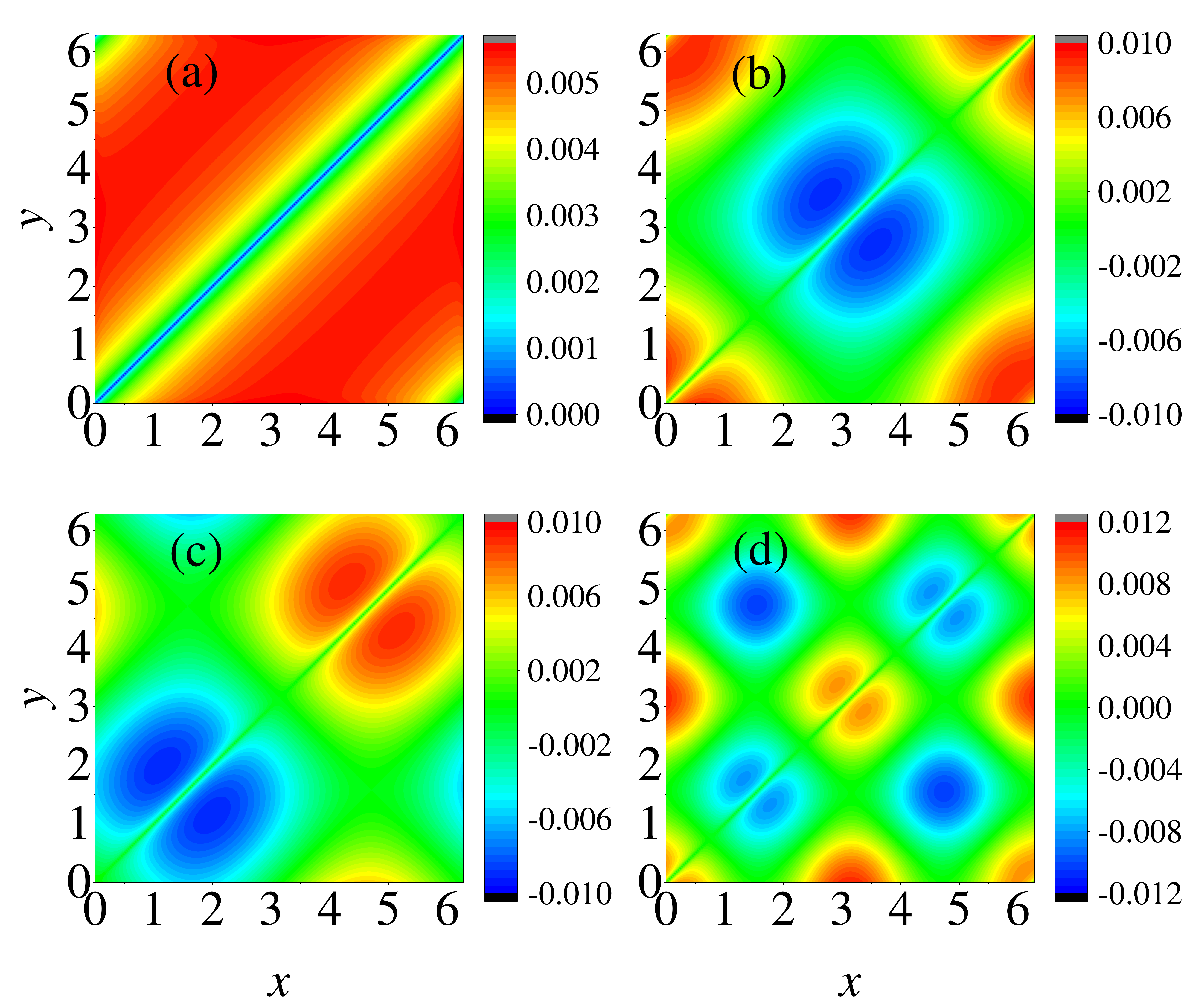}
    \caption{The natural orbital corresponding to the reduced 2-body density matrix $\rho_2(x,y,x^{\prime},y^{\prime})$ for TG gas with $N=51$.  (a) $\phi_0(x,y)$; (b) $\phi_1(x,y)$; (c) $\phi_2(x,y)$; (d) $\phi_3(x,y)$.}
    \label{fig:R2BDM-NO}
\end{figure}
\begin{figure}[htbp]
    \centering
    \includegraphics[width=2.8in]{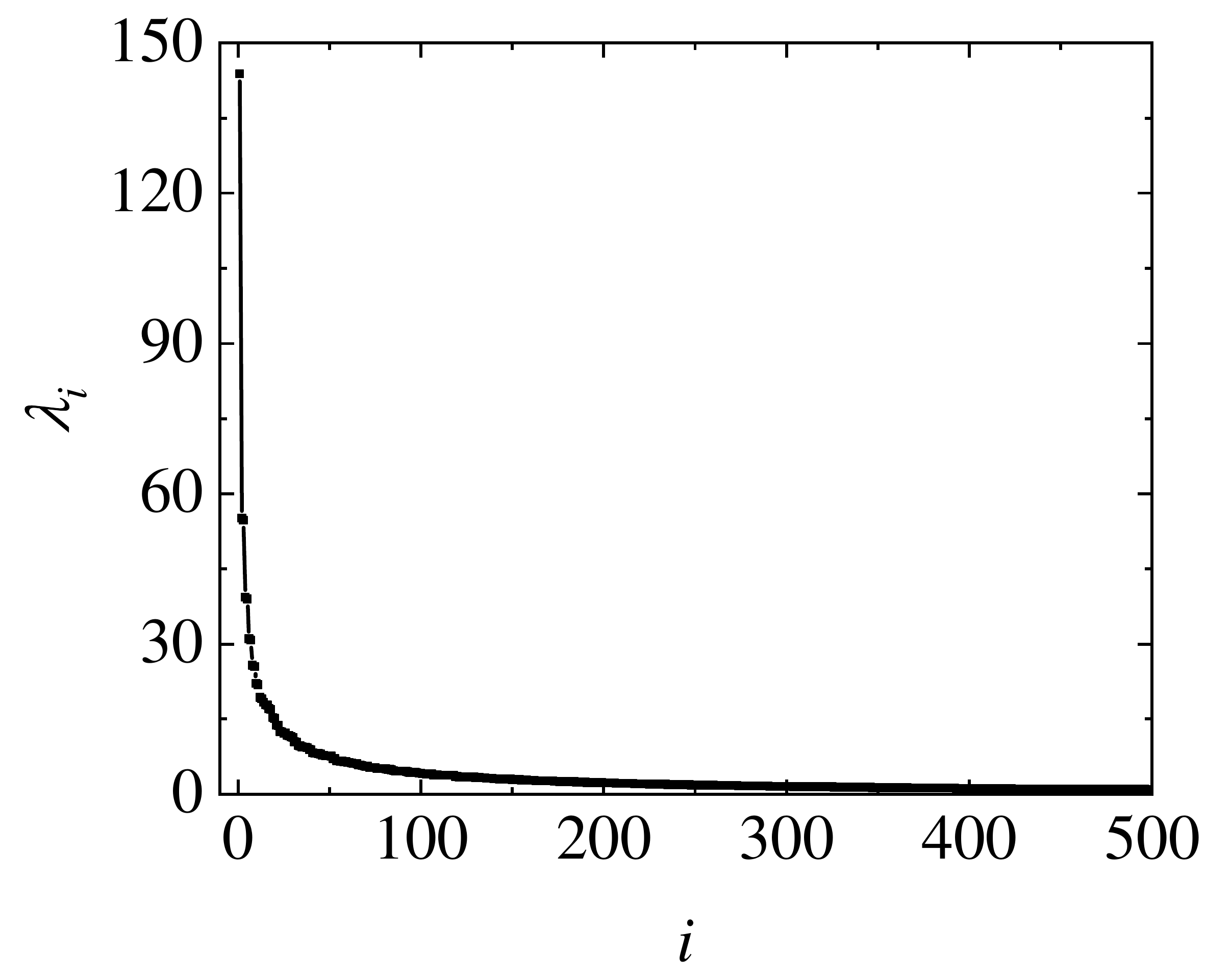}
    \caption{The occupation distribution of natural orbitals $\phi_i(x,y)$ corresponding to the R2BDM $\rho_2(x,y,x^{\prime},y^{\prime})$ for TG gas with $N=51$.}
    \label{fig:OcuNatOrbR2BDM}
\end{figure}
The natural orbitals $\phi_i(x,y)$ and the corresponding occupation numbers $\lambda_i$ can be obtained by solving the following eigen equation
\begin{eqnarray*}
&\int_0^{2\pi} dx^{\prime }dy^{\prime }\rho _{2}(x,y;x^{\prime },y^{\prime })  \phi _i(x^{\prime},y^{\prime})=\lambda_i\phi_i(x,y),
\end{eqnarray*}
where $i$ indexes the natural orbitals and their occupation numbers. The smaller $i$, the larger occupation number. This eigen equation can be solved by replacing the integral with summation by discretizing the coordinate space and the integral equation become the standard algebra eigen equations. The calculation shows that the summation of all occupation numbers satisfy $\sum_i \lambda_i =N(N-1)$, which is consistent with the rigorous result in Ref. \cite{ODLRO}. The natural orbitals $\phi_i(x,y)$ for the four largest occupation number are plotted in Fig. 3. It is shown that natural orbitals are symmetry about $y=x$ and satisfy the periodical boundary condition both in $x$ direction and in $y$ direction. The lowest natural orbital $\phi_0(x,y)$ is translation invariant and depends only on $|x-y|$. The higher natural orbitals $\phi_i(x,y)$ have more nodes with the increase of $i$. Besides the symmetry about $y=x$, $\phi_1(x,y)$ is symmetry about $y=-x$ and $\phi_2(x,y)$ is anti-symmetry about $y=-x$. $\phi_3(x,y)$ is symmetry about both $y=x$ and $y=-x$ and have more nodes. The occupation number distribution of natural orbitals for R2BDM is displayed in Fig. 4. It is shown that with the increase of index number the occupation number of natural orbitals decrease rapidly.

\begin{figure}
    \centering
    \includegraphics[width=3.2in]{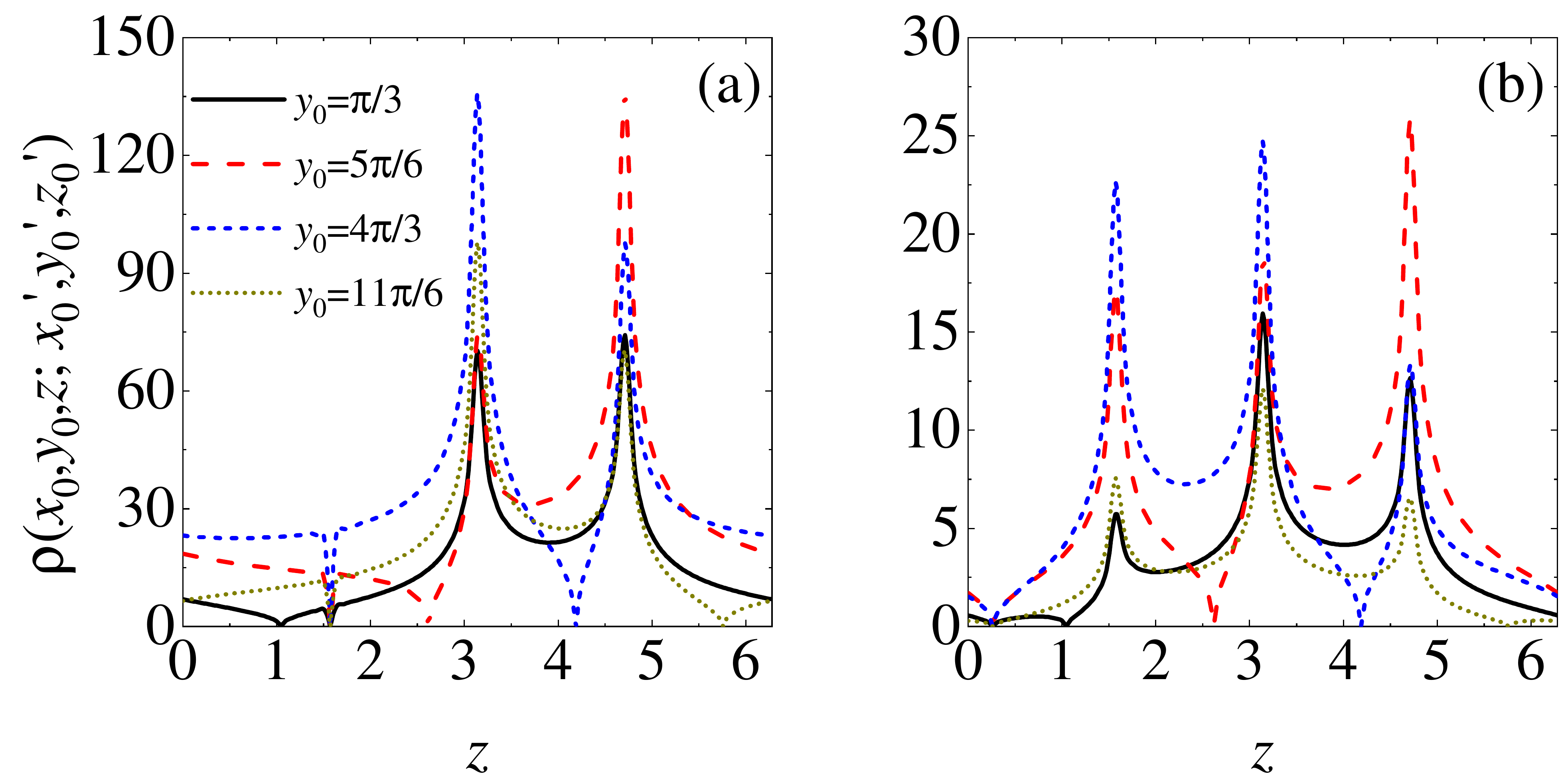}
   \caption{The reduced 3-body density matrix $\rho_3(x_0,y_0,z,x_0^{\prime},y_0^{\prime},z_0^{\prime})$ for TG gas with $N=51$ and $x_0^{\prime}=\pi/2, y_0^{\prime}=\pi, z_0^{\prime}=3\pi/2$. (a) $x_0=\pi/2$; (b) $x_0=\pi/12$.}
    \label{fig:R3BDM-CSz}
\end{figure}
\begin{figure*}
    \centering
    \includegraphics[width=7in]{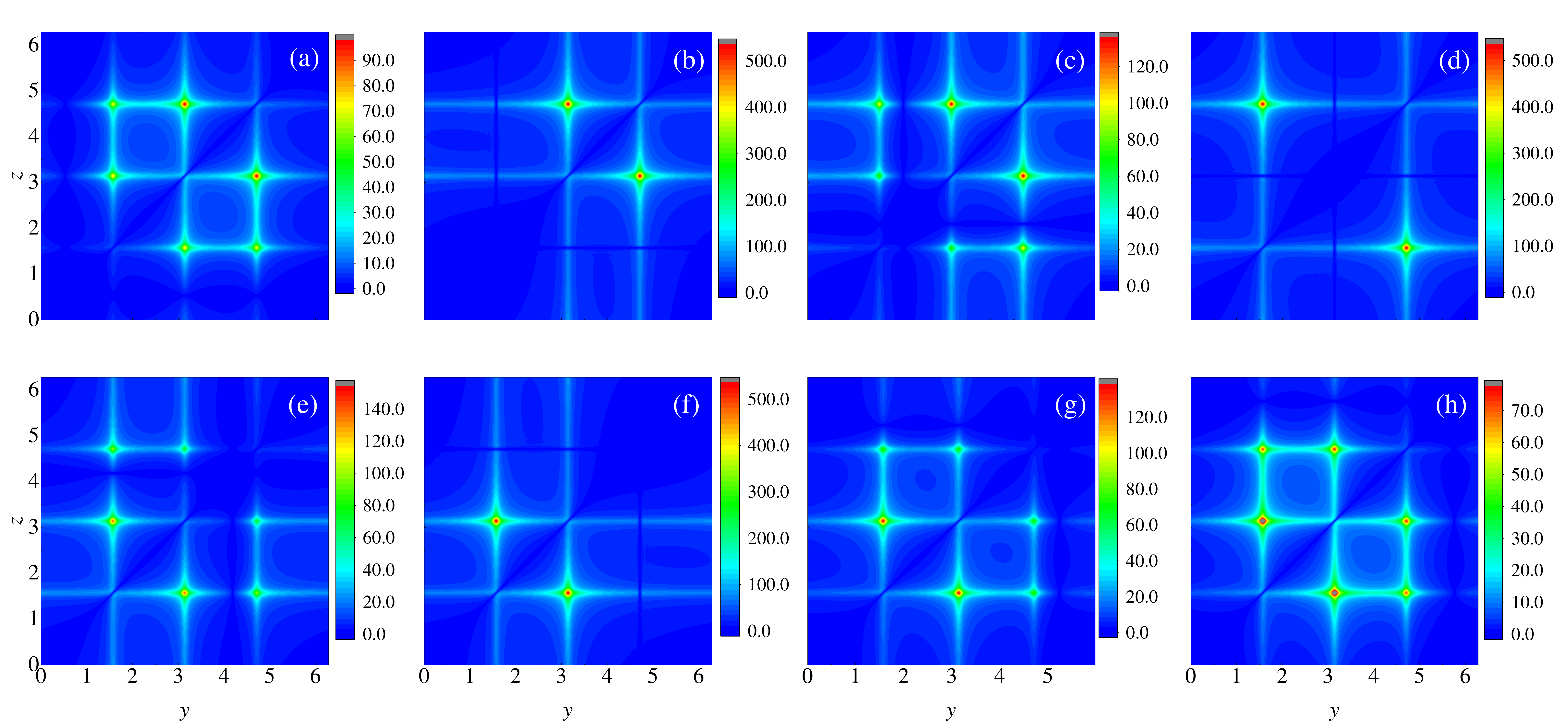}
   \caption{The reduced 3-body density matrix $\rho_3(x_0,y,z,x_0^{\prime},y_0^{\prime},z_0^{\prime})$ for TG gas with $N=51$ and $x_0^{\prime}=\pi/2, y_0^{\prime}=\pi, z_0^{\prime}=3\pi/2$. (a) $x_0=\pi/6$; (b) $x_0=\pi/2$; (c) $x_0=2\pi/3$; (d) $x_0=\pi$; (e) $x_0=4\pi/3$; (f) $x_0=3\pi/2$; (g) $x_0=5\pi/3$; (h) $x_0=11\pi/6$.}
    \label{fig:R3BDM-CS1}
\end{figure*}

The R3BDM is the probability to find three particles at the positions $x^{\prime}$, $y^{\prime}$ and $z^{\prime}$ in the fist measurement and at the positions $x$, $y$ and $z$ in the successive second measurements, which has the following concise form
\begin{eqnarray*}
&&\rho _{3}(x,y,z;x^{\prime },y^{\prime },z^{\prime })   \\
&=&\frac{1}{(2\pi )^{N}}%
e^{i(N-1)(x+y+z-x^{\prime }-y^{\prime }-z^{\prime })/2}\epsilon (x-y)(e^{-ix}-e^{-iy}) \\
&&\times \epsilon (x-z)(e^{-ix}-e^{-iz})\epsilon(y-z)(e^{-iy}-e^{-iz})\epsilon (x^{\prime }-y^{\prime }) \\
&&\times (e^{ix^{\prime }}-e^{iy^{\prime
}})\epsilon (x^{\prime }-z^{\prime })(e^{ix^{\prime }}-e^{iz^{\prime
}})\epsilon (y^{\prime }-z^{\prime })(e^{iy^{\prime }}-e^{iz^{\prime }}) \\
&& \times \det [b_{j-k}(x,y,z;x^{\prime },y^{\prime },z^{\prime
})]_{j,k=1,...,N-3},
\end{eqnarray*}%
where the Toeplitz matrix element $b_{j-k}(x,y,z;x^{\prime },y^{\prime },z^{\prime })$ can be expressed as (here we sort $x,y,z,x^{\prime },y^{\prime },z^{\prime }$ in the ascending order and denote them as $r_i$ ($i=1,2,\cdots,6$) with $r_1 \leq r_2 \leq r_3 \leq r_4 \leq r_5 \leq r_6$)
\begin{eqnarray*}
&&b_{j-k}(x,y,z;x^{\prime },y^{\prime },z^{\prime }) \\
&=&e^{-i(x+y+z)}\sum_{l=0}^{6}(-1)^{l+3}c_{l} \left[ \nu_{j-k-l+3}-2\left( \mu _{j-k-l+3}(r_{1},r_{2}) \right. \right.  \\
&& \left.\left. +\mu_{j-k-l+3}(r_{3},r_{4})+\mu _{j-k-l+3}(r_{5},r_{6})\right) \right] 
\end{eqnarray*}
where the formula of the coefficients $c_l$ ($l=0,1,\cdots,6$) are given in the Appendix A.

In Fig. 5 we show the occupation probability of the third atom in different regimes as we assume that two atoms appear at given $x_0$ and $y_0$ in the successive measurement and that three atoms appear at given atom positions $x_0^{\prime}$, $y_0^{\prime}$ and $z_0^{\prime}$ in the first measurement. We plot $\rho_3(x_0,y_0,z,x_0^{\prime},y_0^{\prime},z_0^{\prime})$ versus $z$ for different $x_0$ and $y_0$ with $x_0^{\prime}=\pi/2$, $y_0^{\prime}=\pi$, $z_0^{\prime}=3\pi/2$, which means the probability that in the successive second measurement one atom appear at $x_0$, the other atom appear at $y_0$, and another atom appear at $z$. Here we assume that in the first measurement three atoms locate at $\pi/2$, $\pi$, and $3\pi/2$, respectively. It is shown that in the second measurement if two atom appear at $x_0$ and $y_0$, the third atom will not appear at their neighbour regions but appear at other regions close to $x_0^{\prime}$, $y_0^{\prime}$, and $z_0^{\prime}$ with larger probability. In the figures zero point and peak means the zero probability and maximum probability, respectively. There are two peaks and two zero points in each lines in Fig. 5a. $\rho_3(x_0,y_0,z;x_0^{\prime},y_0^{\prime},z_0^{\prime})$ are always zero at $z=\pi/2$ (because $x_0=\pi/2$) and $z=y_0$ and it is impossible that one more atom appear at $x_0$ or $y_0$. The black line in Fig. 5a shows that once atoms have populated at $\pi/2$ and $\pi/3$ the third atom will not appear at these two regions but appear at $z=\pi$ and $z=3\pi/2$ with larger probability. The red-dashed lines in Fig. 5a shows that the third atom appear at $z=\pi$ and $z=3\pi/2$ with larger probability and the probability to appear at $\pi/2$ and $5\pi/6$ is zero. In Fig. 5b ($x_0=\pi/12$) $\rho_3(x_0,y_0,z;x_0^{\prime},y_0^{\prime},z_0^{\prime})=0$ for $z=\pi/12$ and $z=y_0$. Since two atoms populate in the regions away from $\pi/2$, $\pi$, and $3\pi/2$, the third atom appears at theses regions with the largest probability. Therefore there are three peaks and two zero points in each lines in Fig. 5b.

In Fig. 6 and in Fig. 7 we show the occupation probability in different regions in the successive measurement when we assume that one atom has occupied the position $x_0$ in the successive second measurement and that three atoms appear at given atom positions $x_0^{\prime}$, $y_0^{\prime}$ and $z_0^{\prime}$ in the first measurement. The R3BDM $\rho_3(x_0,y,z;x_0^{\prime},y_0^{\prime},z_0^{\prime})$ of TG gases for $N=51$ are displayed with $x_0^{\prime}=\pi/2$, $y_0^{\prime}=\pi$ and $z_0^{\prime}=3\pi/2$ for different $x_0$. It is shown that the properties of R3BDM are similar to those of R2BDM qualitatively. Firstly, in the second measurement the probability to two atom appearing at same position is zero, which is shown for all cases. Secondly, in the successive second measurement atoms populate in the neighbour region of $x_0^{\prime}$ $y_0^{\prime}$ and $z_0^{\prime}$ with larger probability. For $x_0=\pi/6$ (Fig. 6a) and $11\pi/6$ (Fig. 6h), $\rho_3(x_0,y,z;x_0^{\prime},y_0^{\prime},z_0^{\prime})$ have the maximum value at the region close to $(y,z)$=$(\pi/2,\pi)$, $(\pi/2,3/\pi)$, $(\pi,\pi/2)$, $(\pi,3\pi/2)$, $(3\pi/2,\pi/2)$, and $(3\pi/2,\pi)$. For $x_0=\pi/2$, $\pi$, or $3\pi/2$, $\rho_3(x_0,y,z;x_0^{\prime},y_0^{\prime},z_0^{\prime})$ becomes zero at the region of $y=x_0$ or $z=x_0$ (Fig. 6b, Fig. 6d, and Fig. 6f). Besides the above regions,  atoms also populate in the regions along $y$=$\pi/2$, $\pi$, and $3\pi/2$ and in the regions along $z$=$\pi/2$, $\pi$, and $3\pi/2$ with a certain probability except $y=x_0$ or $z=x_0$.

\begin{figure}
    \centering
    \includegraphics[width=3.5in]{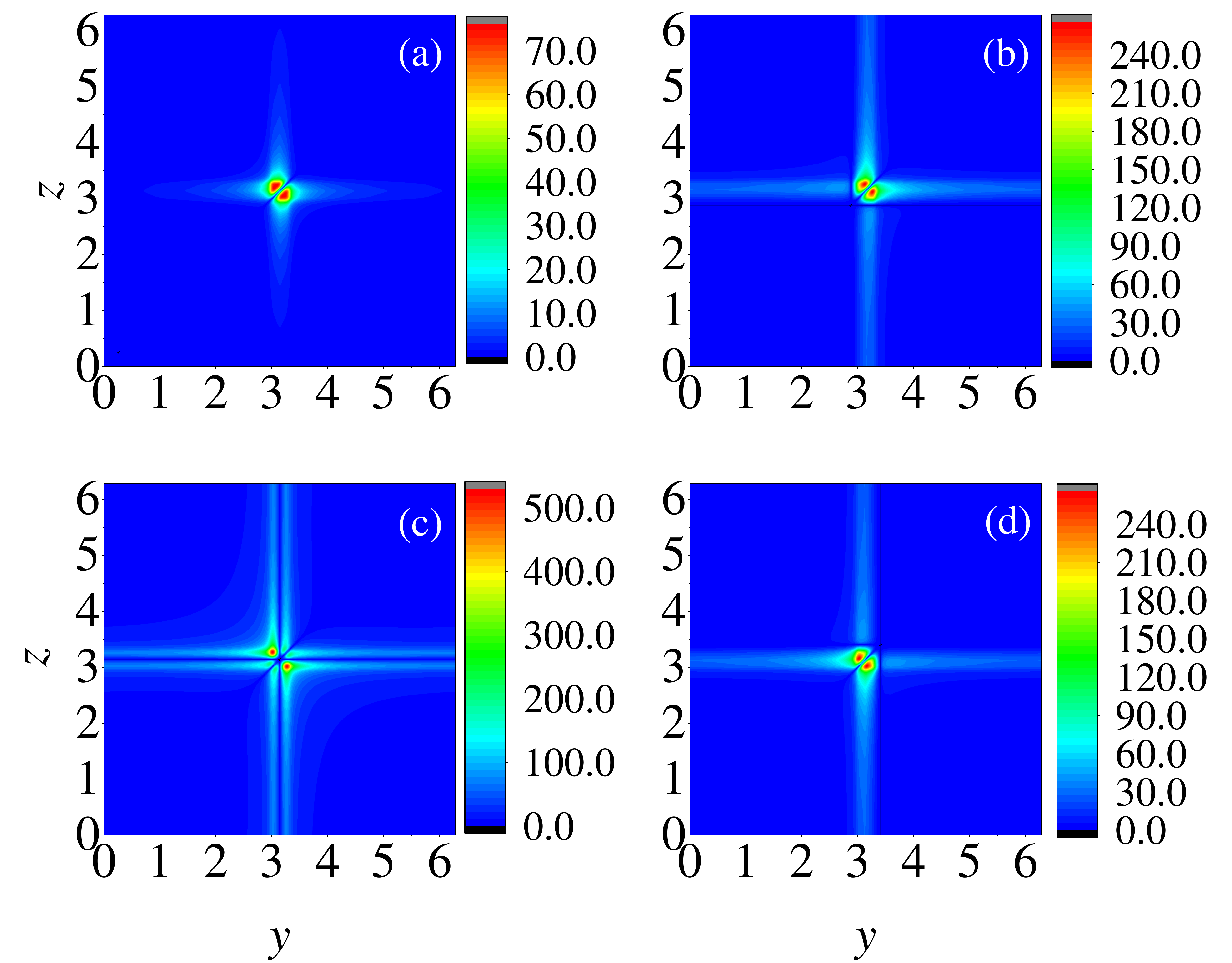}
   \caption{The reduced 3-body density matrix $\rho_3(x_0,y,z,x_0^{\prime},y_0^{\prime},z_0^{\prime})$ for TG gas with $N=51$ and $x_0^{\prime}=23\pi/24, y_0^{\prime}=\pi, z_0^{\prime}=25\pi/24$. (a) $x_0=\pi/12$; (b) $x_0=11\pi/12$; (c) $x_0=\pi$; (d) $x_0=13\pi/12$.}
    \label{fig:R3BDM-CS2}
\end{figure}

In Fig. 7 we plot R3BDM $\rho_3(x_0,y,z,x_0^{\prime},y_0^{\prime},z_0^{\prime})$ for TG gas with $N=51$ and $x_0^{\prime}=23\pi/24, y_0^{\prime}=\pi, z_0^{\prime}=25\pi/24$, i.e., in the first measurement three atoms appear at the neighbour region. It is displayed that in the successive second measurement three atoms still appear at the same regions with the maximum probability. There are finite probability to appear in the region along $y=x_0^{\prime}$, $y_0^{\prime}$, $z_0^{\prime}$ and $z=x_0^{\prime}$, $y_0^{\prime}$, $z_0^{\prime}$ except the region close to $x_0$. Qualitatively the properties of the occupation probability is similar to the situation in Fig. 6 because the constraint of the infinite repulsive contact interactions plays the role only in the nearest neighbour regimes of atoms.



\section{Summary}

In this article, we derived the $n$-body correlation for the ground state of 1D strongly interacting TG gases. Basing on its exact ground state wavefunction obtained by using Bose-Fermi mapping method, the $n$-body correlation functions can be calculated. Using the properties of Vandermonde determinant and the Toeplitz determinant, the $n$-body correlation are formulated explicitly in the form of Toeplitz matrix elements. By further simplifying the integral over multiple particle freedoms, the calculation of $n$-body correlation become feasible. As applications, we study the R2BDM, the corresponding natural orbitals and the occupation distribution. The R3BDM is also investigated.

It is shown that $\rho _{n}(x_{N-n+1},\cdots ,x_{N};x_{N-n+1}^{\prime },\cdots ,x_{N}^{\prime })$ have maximum values in the regions $\{x_{N-n+1},\cdots ,x_{N}\}$= $\{x_{N-n+1}^{\prime },\cdots ,x_{N}^{\prime }\}$ and are zero at $x_j=x_l$ or $x_j^{\prime}=x_l^{\prime}$ ($j,l=N-n+1,\cdots,N$ and $j\neq l$). In conclusion, if in the first measurement atoms appear at $x_{N-n+1}^{\prime}$, $\cdots$, and ,$x_{N}^{\prime }$, respectively, then in the successive second measurement atoms will appear in their neighbour region with the largest probability. In the same measurement the probability for two atom to appear at the same position must be zero.

\begin{acknowledgments}
This work was supported by NSF of China under Grants No. 11774026. L.W. is supported by the National Natural Science Foundation of China (Grant Nos. 11404199, 12147215) and the Natural Science Foundation of Shanxi Province, China (Grant Nos. 2015021012, 1331KSC).
\end{acknowledgments}

\appendix
\section{}
The coefficients of the Toeplitz matrix element of the R3BDM:
\begin{widetext}
\begin{eqnarray*}
c_0&=&1,    \\
c_1&=&e^{ix}+e^{iy}+e^{iz}+e^{ix^{\prime }}+e^{iy^{\prime
}}+e^{iz^{\prime }},  \\
c_2&=&e^{i(x+y)}+e^{i(x+z)}+e^{i(x+x^{\prime})}+e^{i(x+y^{\prime
})}+e^{i(x+z^{\prime })}+e^{i(y+z)}+e^{i(y+x^{\prime})}+e^{i(y+y^{\prime
})}+e^{i(y+z^{\prime })}  \\
&& +e^{i(z+x^{\prime })}+e^{i(z+y^{\prime })}+e^{i(z+z^{\prime
})}+e^{i(x^{\prime }+y^{\prime })}+e^{i(x^{\prime }+z^{\prime
})}+e^{i(y^{\prime }+z^{\prime })}, \\
c_3 &=& e^{i(x+y+z)}+e^{i(x+y+x^{\prime })}+e^{i(x+y+y^{\prime
})}+e^{i(x+y+z^{\prime })}+e^{i(x+z+x^{\prime })}+e^{i(x+z+y^{\prime
})}+e^{i(x+z+z^{\prime })}+e^{i(x+x^{\prime }+y^{\prime})} \\
&&+e^{i(x+x^{\prime}+z^{\prime })}+e^{i(x+y^{\prime }+z^{\prime })}+e^{i(y+z+x^{\prime })}+e^{i(y+z+y^{\prime})}+e^{i(y+z+z^{\prime})}+e^{i(y+x^{\prime }+y^{\prime })}+e^{i(y+x^{\prime }+z^{\prime
})}+e^{i(y+y^{\prime }+z^{\prime })}  \\
&& +e^{i(z+x^{\prime }+y^{\prime
})}+e^{i(z+x^{\prime }+z^{\prime })}+e^{i(z+y^{\prime }+z^{\prime
})}+e^{i(x^{\prime }+y^{\prime }+z^{\prime })}, \\
c_4 &=& e^{i(x+y+z+x^{\prime })}+e^{i(x+y+z+y^{\prime
})}+e^{i(x+y+z+z^{\prime })}+e^{i(x+y+x^{\prime }+y^{\prime
})}+e^{i(x+y+x^{\prime }+z^{\prime })}+e^{i(x+y+y^{\prime }+z^{\prime
})}\\
&&+e^{i(x+z+x^{\prime }+y^{\prime })}+e^{i(x+z+x^{\prime }+z^{\prime
})} +e^{i(x+z+y^{\prime }+z^{\prime })}+e^{i(x+x^{\prime }+y^{\prime
}+z^{\prime })}+e^{i(y+z+x^{\prime }+y^{\prime })}+e^{i(y+z+x^{\prime
}+z^{\prime })} \\
&& +e^{i(y+z+y^{\prime }+z^{\prime })}+e^{i(y+x^{\prime
}+y^{\prime }+z^{\prime })}+e^{i(z+x^{\prime }+y^{\prime }+z^{\prime
})},  \\
c_5 &=& e^{i(x+y+z+x^{\prime }+y^{\prime })}+e^{i(x+y+z+x^{\prime
}+z^{\prime })}+e^{i(x+y+z+y^{\prime }+z^{\prime })}+e^{i(x+y+x^{\prime
}+y^{\prime }+z^{\prime })}+e^{i(x+z+x^{\prime }+y^{\prime }+z^{\prime
})}+e^{i(y+z+x^{\prime }+y^{\prime }+z^{\prime })},  \\
c_6&=&e^{i(x+y+z+x^{\prime }+y^{\prime }+z^{\prime })}.
\end{eqnarray*}
\end{widetext}


\begin{thebibliography}{99}

\bibitem{Glauber} R. Glauber, Phys. Rev. {\bf 130}, 2529 (1963).

\bibitem{ABergschneider2019} A. Bergschneider, V. M. Klinkhamer, J. H. Becher, R. Klemt, L. Palm, G. Z{\" u}rn, S. Jochim, and P. M. Preiss, Nat. Phys. {\bf 15}, 640 (2019).

\bibitem{HBT} R. Hanbury Brown and R. Q. Twiss, Nature {\bf 178}, 1046 (1956).

\bibitem{MSchellekens} M. Schellekens, R. Hoppeler, A. Perrin, J. V. Gomes, D. Boiron, A. Aspect, and C. I. Westbrook, Science {\bf 310}, 648 (2005).

\bibitem{TJeltes} T. Jeltes et al., Nature {\bf 445}, 402 (2007).

\bibitem{SSHodgman2011} S. S. Hodgman, R. G. Dall, A. G. Manning, K. G. H. Baldwin, and A. G. Truscott, Science {\bf 331}, 1046 (2011).

\bibitem{PMPreiss} P. M. Preiss, J. H. Becher, R. Klemt, V. Klinkhamer, A. Bergschneider, N. Defenu, and S. Jochim, Phys. Rev. Lett. {\bf 122}, 143602 (2019).

\bibitem{VGuarrera} V. Guarrera, P. Wurtz, A. Ewerbeck, A. Vogler, G. Barontini, and H. Ott, Phys. Rev. Lett. \textbf{107}, 160403 (2011).

\bibitem{RGDall} R. G. Dall, A. G. Manning, S. S. Hodgman, W. RuGway, K. V. Kheruntsyan, and A. G. Truscott, Nature Physics {\bf 9}, 341 (2013).

\bibitem{GHerce} G. Herc{\'e}, J.-P. Bureik, A. T{\'e}nart, A. Aspect, A. Dareau, and D. Cl{\'e}ment, arXiv.2207.14070.

\bibitem{ABergschneider2018} A. Bergschneider, V. M. Klinkhamer, J. H. Becher, R. Klemt, G. Z{\" u}urn, P. M. Preiss, and S. Jochim, Phys. Rev. A {\bf 97}, 063613 (2018).

\bibitem{MHolten} M. Holten, L. Bayha, K. Subramanian, S. Brandstetter, C. Heintze, P. Lunt, P. M. Preiss, and S. Jochim, Nature {\bf 606}, 287 (2022).

\bibitem{SSHodgman2017} S. S. Hodgman, R. I. Khakimov, R. J. Lewis-Swan, A. G. Truscott, and K. V. Kheruntsyan, Phys. Rev. Lett. {\bf 118}, 240402 (2017).

\bibitem{SSHodgman2019} S. S. Hodgman, W. Bu, S. B. Mann, R. I. Khakimov, and A. G. Truscott, Phys. Rev. Lett. {\bf 122}, 233601 (2019).

\bibitem{NJVDruten}  N. J. van Druten and W. Ketterle, Phys. Rev. Lett. {\bf 79}, 549 (1997).

\bibitem{BParedes}  B. Paredes, A. Widera, V. Murg, O. Mandel, S. F\"{o}lling, I. Cirac, G. V. Shlyapnikov, T. W. H\"{a}nsch, and I. Bloch,
Nature {\bf 429}, 277 (2004).

\bibitem{TKinoshita}  T. Kinoshita, T. Wenger and D. S. Weiss, Science {\bf 305}, 1125 (2004).

\bibitem{TJacqmin} T. Jacqmin, J. Armijo, T. Berrada, K. V. Kheruntsyan, and I. Bouchoule, Phys. Rev. Lett. \textbf{106}, 230405 (2011).

\bibitem{LTonks} L. Tonks, Phys. Rev. \textbf{50}, 955 (1936).

\bibitem{MDGirardeau1960} M. D. Girardeau, J. Math. Phys. \textbf{1}, 516 (1960).

\bibitem{RMP2011} M. A. Cazalilla, R. Citro, T. Giamarchi, E. Orignac, M. Rigol, Rev. Mod. Phys. \textbf{83}, 1405 (2011).

\bibitem{RMP2012} A. Imambekov, T. L. Schmidt, and L. I. Glazman, Rev. Mod. Phys. \textbf{84}, 1253 (2012).

\bibitem{RMP2013} X.-W. Guan, M. T. Batchelor, and C. Lee,  Rev. Mod. Phys. \textbf{85}, 1633 (2013).

\bibitem{EJKPN} EJKP Nandani and X.-W. Guan 2018 Chinese Phys. B \textbf{27} 070306 (2018).

\bibitem{NFabbri} N. Fabbri, M. Panfil, D. Cl\'{e}ment, L. Fallani, M. Inguscio, C. Fort, and J.-S. Caux, Phys. Rev. A \textbf{91}, 043617 (2015).

\bibitem{WXu} W. Xu and M. Rigol, Phys. Rev. A \textbf{92}, 063623 (2015).

\bibitem{VGritsev} V. Gritsev, E. Altman, E. Demler, and A. Polkovnikov, Nat. Phys. \textbf{2}, 705 (2006).

\bibitem{LMathey} L. Mathey, A. Vishwanath, and E. Altman, Phys. Rev. A \textbf{79}, 013609 (2009).

\bibitem{KHe} K. He and M. Rigol, Phys. Rev. A \textbf{83}, 023611 (2011).

\bibitem{PDevillard2020} P. Devillard, D. Chevallier, P. Vignolo, and M. Albert, Phys. Rev. A  \textbf{101}, 063604 (2020).

\bibitem{ATenart} A. Tenart, G. Herce, J. P. Bureik, A. Dareau, and D. Clement, Nat. Phys. {\bf 17}, 1364 (2021).

\bibitem{Olshanii2017} M. Olshanii, V. Dunjko, A. Minguzzi, and G. Lang, Phys. Rev. A \textbf{96}, 033624 (2017).

\bibitem{PDevillard2021} P. Devillard, A. Benzahi, P. Vignolo, and M. Albert, Phys. Rev. A \textbf{104}, 053306 (2021).

\bibitem{ALenard} A. Lenard, J. Math. Phys. {\bf 5}, 930 (1964).

\bibitem{PJForrester} P. J. Forrester, N. E. Frankel, T. M. Garoni, and N. S. Witte, Phys. Rev. A {\bf 67}, 043607 (2003).

\bibitem{ILovas2017A} I. Lovas, B. Dora, E. Demler, and G. Zarand, Phys. Rev. A \textbf{95}, 053621  (2017).

\bibitem{ILovas2017B} I. Lovas, B. Dora, E. Demler, and G. Zarand, Phys. Rev. A \textbf{95}, 023625  (2017).

\bibitem{VIYukalov} V. I. Yukalov and M. D. Girardeau, Laser Phys. Lett. \textbf{2}, 375 (2005).

\bibitem{RPezer} R. Pezer and H. Buljan, Phys. Rev. Lett. \textbf{98}, 240403 (2007).

\bibitem{JDobrzyniecki} J. Dobrzyniecki and T. Sowinski, Phys. Rev. A \textbf{99}, 063608 (2019).

\bibitem{TPapenbrock} T. Papenbrock, Phys. Rev. A \textbf{67}, 041601 (R) (2003).

\bibitem{HGVaidyaJMP} H. G. Vaidya and C. A. Tracy, J. Math. Phys. \textbf{20}, 2291 (1979).

\bibitem{DMGangardt} D. M. Gangardt, J. Phys. A \textbf{37}, 9335 (2004).

\bibitem{MDGirardeau2001} M. D. Girardeau, E. M. Wright, and J. M. Triscari, Phys. Rev. A {\bf 63}, 033601 (2001).

\bibitem{ODLRO} C. N. Yang, Rev. Mod. Phys. {\bf 34}, 694 (1962).

\end{thebibliography}
\end{document}